\documentclass[a4paper,fleqn]{cas-dc}

\usepackage[numbers,sort&compress]{natbib}
\usepackage{amsmath,amssymb}
\usepackage{algorithm}
\usepackage[noend]{algpseudocode}
\usepackage{subcaption}
\usepackage{booktabs}
\usepackage{pifont}
\usepackage{url}
\usepackage{float}
\usepackage{ragged2e}
\usepackage[final]{microtype}

\floatname{algorithm}{Algorithm}
\algrenewcommand\algorithmicrequire{\textbf{Input:}}
\algrenewcommand\algorithmicensure{\textbf{Output:}}
\AtBeginEnvironment{algorithmic}{\footnotesize\raggedright}
\DeclareMathOperator{\round}{round}

\ExplSyntaxOn
\cs_set:Npn \__cas_head: {}
\ExplSyntaxOff

\begin{document}
\pagestyle{cas}
\let\WriteBookmarks\relax
\def\floatpagepagefraction{1}
\def\textpagefraction{.001}
\setlength{\emergencystretch}{0.5em}
\setlength{\textfloatsep}{6pt plus 2pt minus 2pt}
\setlength{\floatsep}{6pt plus 2pt minus 2pt}
\setlength{\intextsep}{6pt plus 2pt minus 2pt}
\setlength{\dbltextfloatsep}{8pt plus 2pt minus 2pt}
\setlength{\dblfloatsep}{8pt plus 2pt minus 2pt}

\shorttitle{FlashT2S}
\shortauthors{R. Guo et al.}

\title[mode=title]{FlashT2S: GPU-Accelerated Tensor-to-Spike Conversion Operators for Heterogeneous Neuromorphic LLM Backends}

\author[1]{Rongyuan Guo}
\ead{rongyuanguo@hnu.edu.cn}

\author[1,2]{Haotian Wang}
\cormark[1]
\ead{wanghaotian@hnu.edu.cn}

\author[1]{Yufeng Zhang}
\cormark[1]
\ead{yufengzhang@hnu.edu.cn}

\author[1]{Wangdong Yang}
\ead{yangwangdong@hnu.edu.cn}

\author[1]{Kenli Li}
\ead{lkl@hnu.edu.cn}

\affiliation[1]{
  organization={College of Computer Science and Electronic Engineering, Hunan University},
  city={Changsha},
  country={China}
}

\affiliation[2]{
  organization={Yuelushan Center for Industrial Innovation},
  city={Changsha},
  state={Hunan},
  country={China}
}

\cortext[cor1]{Corresponding author}

\begin{abstract}
Artificial intelligence faces two growing challenges: computational bottlenecks and energy inefficiency inherent to the von Neumann architecture. Neuromorphic computing offers a promising alternative, yet its immature ecosystem still relies on conventional systems for data supply, requiring efficient conversion between tensor and spike dataflow.

We observe a clear \textbf{mismatch}, where data conversion is slow while downstream computation is fast. As data size grows quadratically with large language models, this mismatch becomes a major bottleneck.

In this work, we examine the data-conversion components that appear on the critical path of training and inference in neuromorphic LLMs. The work presents \textbf{FlashT2S}(Flash Tensor-to-Spike): GPU-Accelerated Tensor-to-Spike Conversion Operators, which applies architecture-aware and data-centric optimizations to reduce conversion latency. Experimental results show that FlashT2S delivers an average speedup of \textbf{850.78$\times$} over the native operators across individual operators and \textbf{802.83$\times$} in realistic workloads. Additionally, compared with GPU kernels automatically generated by \texttt{torch.compile}, our kernels achieve \textbf{5--11$\times$} speedups. Notably, it achieves near-optimal absolute performance: large-scale conversions are reduced to tens of microseconds \textbf{(10--99 $\mu$s)}, while small-scale conversions reach single-digit microsecond levels \textbf{(<10 $\mu$s)}. Furthermore, its standardized output enables seamless integration across heterogeneous backends, including diverse neuromorphic hardware.
\end{abstract}


\maketitle
\section{Introduction}
\textbf{Background and Challenge.} Traditional von Neumann architectures face performance and energy bottlenecks due to the ``memory wall''~\cite{liu2024future,gholami2024ai}. While heterogeneous CPU/GPU--neuromorphic systems offer low-power alternatives~\cite{al2024survey,kudithipudi2025neuromorphic}, they suffer from a software gap: inefficient tensor-to-spike dataflow conversion~\cite{hwang2024spikedattention,pan2025spikingbrain}, whose mismatched execution and memory models impose high energy and synchronization overhead~\cite{khadse2025neuromorphic,xu2024optimizing}.

\textbf{Issue.} Following existing prior work on hybrid systems combining CPUs/GPUs with neuromorphic chips \cite{pan2025spikingbrain}, we profile the execution of attention and feed-forward layers. Our results (Figure 1) show that while high-performance libraries like cuBLAS and OpenBLAS significantly reduce matrix multiplication time, the overall latency is increasingly dominated by data conversion. As input size(sequence length) grows, these conversion tasks become the primary bottleneck, severely limiting end-to-end performance.

\begin{figure}[pos=ht]
  \centering
  \includegraphics[width=\columnwidth]{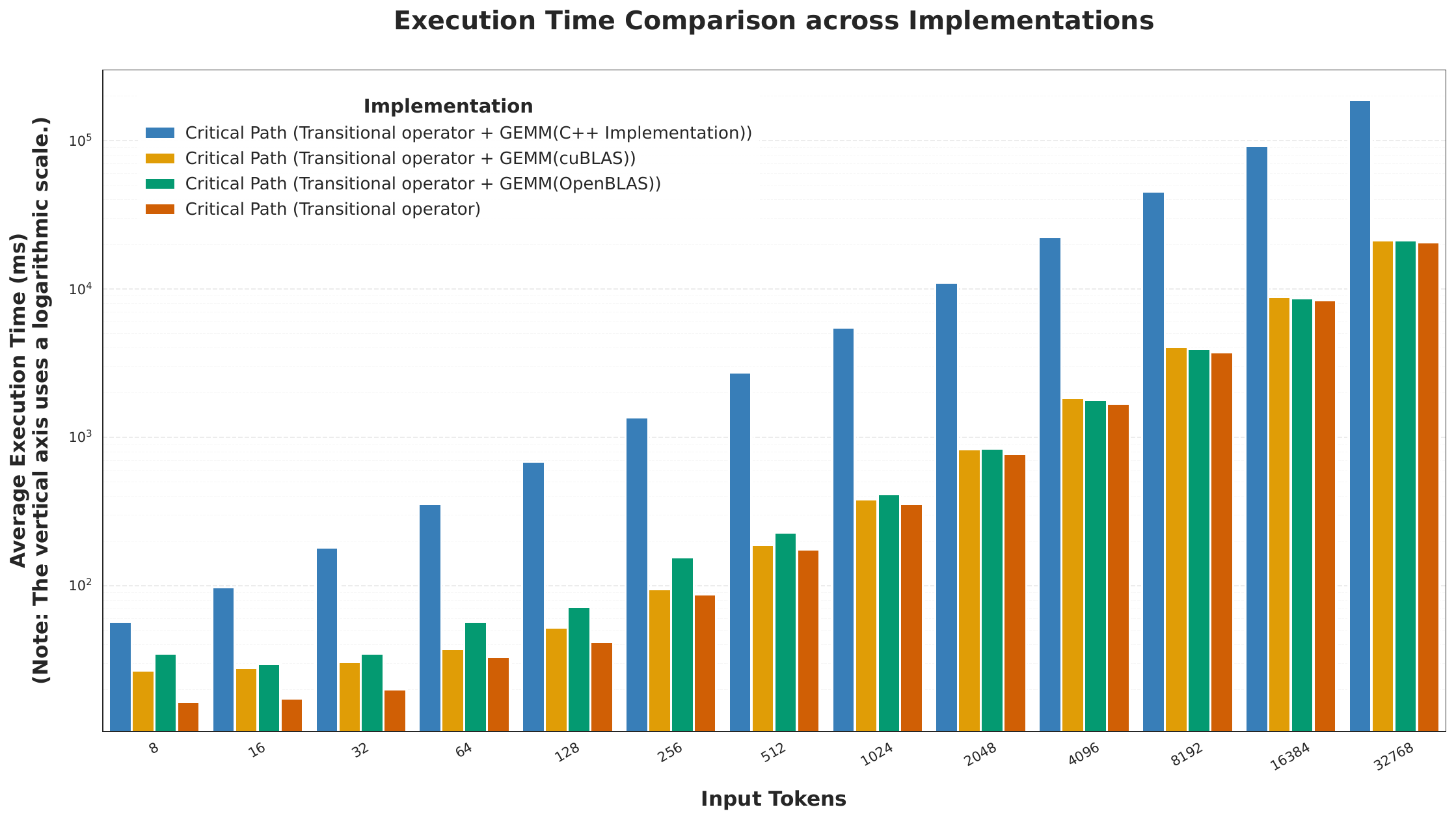}
  \caption{Execution time comparison (log-scale).}
  \label{fig:log_inference_time}
\end{figure}

Meanwhile, we also identify that these data-conversion operators serve as crucial bridges between the two architectures, with broad applicability and transferability. These operators include adaptive thresholding to prevent neuron over-activation and inactivity, event-driven spike encoding, and INT8 quantization tailored for spike characteristics. Therefore, we decided to optimize these operators to break their bottleneck effect on downstream data computation, and endow them with greater independence, transferability, and adaptability.

This work makes the following contributions.

We propose GPU-accelerated parallel algorithms for three critical conversion operators: adaptive thresholding, event-driven spike encoding, and INT8 quantization, exposing fine-grained parallelism to break the performance bottlenecks in tensor-to-spike conversion.

We integrate these optimizations into FlashT2S, a lightweight library that achieves strong performance over native baselines, compiler-generated GPU kernels, and realistic workloads, while reducing conversion latency to the microsecond level.

Our implementation marks a meaningful step toward bridging the latency gap between von Neumann and neuromorphic architectures. Beyond performance gains, this work provides a theoretical and practical framework for the CUDA parallelization of complex, multi-step operators in heterogeneous computing systems.

The source code is publicly available at \url{https://gitee.com/GuoEulerLab/SpikeTransOps} for use by potential users.

\section{Background }
This section details data-conversion operators for hybrid systems. We focus on the co-design of high-performance algorithms and their hardware-efficient implementation to bridge the gap between continuous tensors and sparse spikes.

\subsection{Operator System Distribution}
\begin{figure}[pos=ht]
  \centering
  \includegraphics[width=0.95\columnwidth]{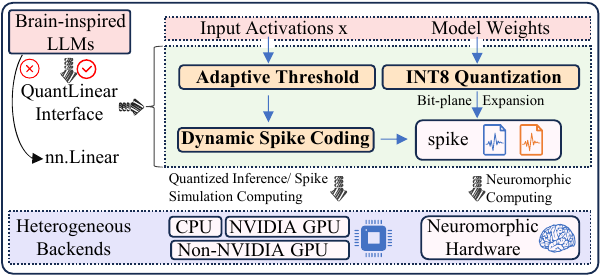}
  \caption{Operator distribution in brain-inspired/LLM inference.}
  \label{fig:operator_distribution}
\end{figure}

The proposed operator system bridges brain-inspired LLMs and diverse computing platforms by replacing the standard \texttt{nn.Linear} with a specialized QuantLinear Interface (Fig.~\ref{fig:operator_distribution}). Within this interface, the Adaptive Threshold, INT8 Quantization, and Dynamic Spike Coding operators are responsible for efficient data conversion across heterogeneous hardware.

\subsection{Data Transformation Operators}
Motivated by the challenges and issues identified in Section 1, this work focuses on three key operators for converting dense CPU/GPU tensor flows into sparse spike flows:

\noindent\textbf{(1) Adaptive Thresholding.} 
To prevent neuron over-activation, the threshold $V_{th}$ is dynamically computed from input $x$: $V_{th}(x) = \frac{1}{k} \cdot \text{mean}(|x|)$. The integer spike count $s_{INT}$ is then generated via $s_{INT} = \text{round}(x / V_{th}(x))$.

\noindent\textbf{(2)Bitwise Spike Encoding.} 
This operator converts $s_{INT}$ into sequences $s_t$ using three modes:

\noindent\ding{172} \textbf{Mode 1 (Unidirectional):}
$s_t = (s_{INT} \gg (N-1-t)) \& 1$ for non-negative.

\noindent\ding{173} \textbf{Mode 2 (Bidirectional Non-Complement):}
$s_t = \text{sign}(s_{INT}) \cdot [(|s_{INT}| \gg (N-1-t)) \& 1]$.

\noindent\ding{174} \textbf{Mode 3 (Bidirectional Two's Complement):}
Standard complement representation for $s_{INT} < 0$.

To optimize efficiency, two shift strategies are used: \textit{Half-Range Compression} ($s_{INT}' = s_{INT} - \lceil s_{INT}/2 \rceil$) for non-negative inputs, and \textit{Baseline Translation} ($s_{INT}' = s_{INT} - \min(s_{INT})$) to avoid negatives. The bit-width $N$ is dynamically set as $\lceil \log_2(\max(|s_{INT}|)+1) \rceil$ (plus 1 for Mode 3).

\noindent\textbf{(3) INT8 Group Quantization.} 
Weights $W$ are split into $G$ groups. Each group $g$ is quantized as
\begin{equation}
W_{\text{int8}}^{(g)} =
\operatorname{clip}\!\left(
\operatorname{round}\!\left(W^{(g)} / s_W^{(g)}\right), -128, 127
\right),
\end{equation}
where the scaling factor is $s_W^{(g)} = \max(|W^{(g)}|) / 127$.

\section{FlashT2S: Design and Implementation}

This section presents the design of FlashT2S, a suite of GPU-accelerated tensor-to-spike conversion operators for heterogeneous neuromorphic LLM backends. FlashT2S is engineered to support heterogeneous systems integrating CPUs, GPUs, and neuromorphic accelerators for efficient LLM inference. We focus on three critical components: adaptive thresholding, bitwise spike encoding, and grouped weight quantization.

\subsection{Framework Design}
The FlashT2S framework integrates three specialized data-conversion pipelines bridging continuous tensors and discrete spikes with microsecond-level latency.

\begin{figure}[pos=ht]
  \centering
  \includegraphics[width=\columnwidth]{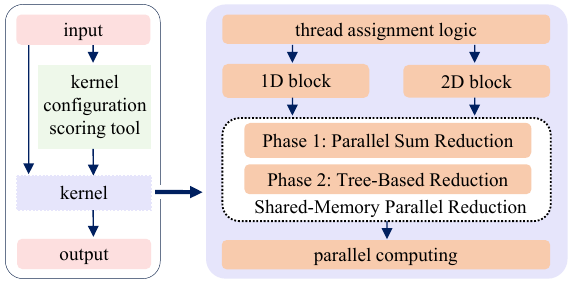}
  \caption{Adaptive Threshold Framework.}
  \label{fig:framework_threshold}
\end{figure}

\textbf{GPU-Based Adaptive Threshold Computing Framework.} 
As illustrated in Fig.~\ref{fig:framework_threshold}, this framework provides a GPU-native acceleration scheme for calculating neuron-specific thresholds dynamically. It begins with an offline \textit{Kernel Configuration Tool} that determines the optimal thread-block layout based on input dimensions to maximize occupancy. During runtime, the system dynamically allocates 1D or 2D blocks to handle varying sequence length(tokens). The core computation is structured as a high-throughput pipeline: Stage 1 executes parallel summation across the embedding dimension to gather per-token statistics, followed by Stage 2 which employs a shared-memory tree reduction. This design ensures that threshold updates remain on the critical path with minimal latency, even for large-scale sequence length.

\begin{figure}[pos=ht]
  \centering
  \includegraphics[width=\columnwidth]{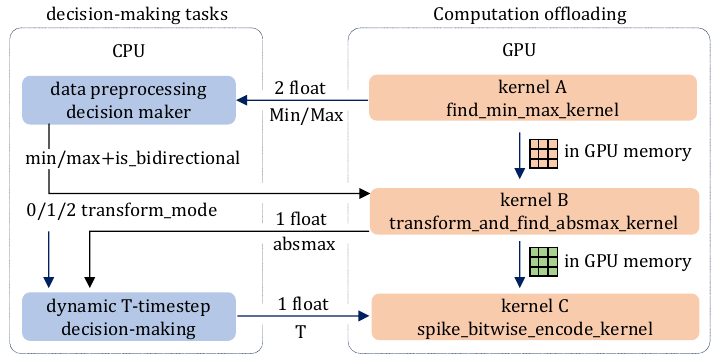}
  \caption{Spike Encoding Dataflow.}
  \label{fig:framework_encoding}
\end{figure}

\textbf{CPU-GPU Collaborative Spike Encoding Architecture.} 
To handle the complexity of bitwise spike coding without the heavy synchronization overhead typical of traditional methods, this architecture (Fig.~\ref{fig:framework_encoding}) partitions the task into three GPU kernels interleaved with lightweight CPU logic. \textit{Kernel A} first computes global extremes ($x_{\min}, x_{\max}$) via parallel reduction, returning two floats for the CPU to select the optimal coordinate transformation (i.e., Half-Range Compression or Baseline Translation). \textit{Kernel B} then performs in-place coordinate transformation and returns the absolute maximum for CPU-side bit-width $T$ calculation. Finally, \textit{Kernel C} executes parallel bitwise expansion into sparse spike sequences. By limiting Host-Device communication to four floats, this design bypasses PCIe bottlenecks, achieving near-microsecond latency.

\begin{figure}[pos=ht]
  \centering
  \includegraphics[width=\columnwidth]{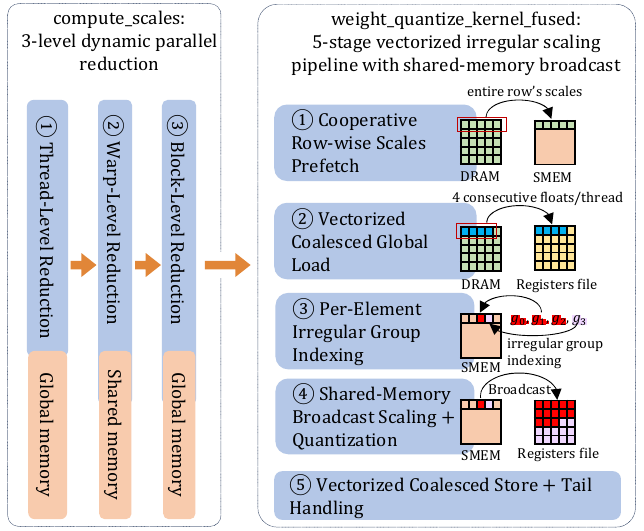}
  \caption{Weight Quantization Pipeline.}
  \label{fig:framework_quant}
\end{figure}

\textbf{Hierarchical Grouped Weight Quantization Pipeline.} 
The weight quantization framework, depicted in Fig.~\ref{fig:framework_quant}, implements a standalone operator suite for per-channel grouped Int8 symmetric quantization. The dataflow is organized into a two-stage vectorized pipeline. In the first stage, it computes scaling factors using a three-level hierarchical reduction (thread, warp, and block levels), adapting seamlessly to different group sizes. In the second stage, it performs fused quantization by prefetching scales into fast shared memory and utilizing \texttt{float4} vectorized instructions for memory access. This hierarchical memory management ensures high global memory bandwidth utilization while maintaining numerical precision for neuromorphic LLM inference.

\subsection{Algorithm Design}
Guided by the framework designs presented above, we develop fine-grained parallel algorithms for each operator to maximize the acceleration of multi-step complex computations using CUDA cores.These algorithmic designs are presented below.

\textbf{GPU Parallel Adaptive Thresholding.} Algorithm~\ref{alg:gpu_adaptive_threshold} implements a dynamic thresholding scheme for real-time spike encoding. The process begins with grid configuration (Lines 2--5), where block layouts are adaptively selected based on batch size $B$ to maximize hardware occupancy. The core logic consists of four phases: Phase 1 (Lines 8--11) performs a parallel sum of absolute values using a grid-stride loop (14 elements per thread) to compute the $L_1$ norm; Phase 2 (Lines 12--16) executes an 8-level tree reduction within shared memory to aggregate row-wise statistics; Phase 3 (Lines 17--20) calculates the adaptive threshold $V_{\text{th}}$ using a clamp operation for numerical stability; and Phase 4 (Lines 21--23) completes element-wise quantization. This hierarchical design reduces branch divergence and ensures high-throughput integer spike generation.

\algrenewcommand\algorithmicrequire{\textbf{Input:}}
\algrenewcommand\algorithmicensure{\textbf{Output:}}

\begin{algorithm}[ht]
\caption{GPU Parallel Adaptive Threshold Quantization}
\label{alg:gpu_adaptive_threshold}
\begin{algorithmic}[1]
\Require Tensor $X \in \mathbb{R}^{B \times D}$, $D = 3584$, scaling $k > 0$
\Ensure  Spikes $S \in \mathbb{R}^{B \times D}$, thresholds $V_{\text{th}} \in \mathbb{R}^{B}$
\State \textit{// Grid Configuration}
\If{$B \leq 64$}\Comment{1 row per block}
    \State Grid: $(B, 1)$, Block: $(256, 1)$, Shared: 1KB
\Else            \Comment{4 rows per block}
    \State Grid: $(\lceil B/4 \rceil, 1)$, Block: $(256, 4)$, Shared: 4KB
\EndIf
\ForAll{blocks in parallel}
    \State Compute row index, check bounds, get $\text{tid}$
    \State \textit{// Phase 1: Parallel Sum}
    \For{14 iterations per thread} \Comment{Each thread: 14 elements}
        \State Accumulate $|X_{\text{row},j}|$ with stride 256
    \EndFor
    \State Store to shared memory and synchronization
    \State \textit{// Phase 2: Tree Reduction}
    \For{$s \in \{128, 64, 32, 16, 8, 4, 2, 1\}$}
        \If{$\text{tid} < s$}
            \State $\texttt{sdata}[\text{tid}] \leftarrow \texttt{sdata}[\text{tid}] + \texttt{sdata}[\text{tid} + s]$
        \EndIf
        \State Synchronize threads
    \EndFor
    \State \textit{// Compute Threshold}
    \If{$\text{tid} = 0$}\Comment{Single thread per row}
        \State $V_{\text{th}}[\text{row}] \leftarrow \operatorname{clamp}(\texttt{sdata}[0] / (D \times k), 10^{-5}, 10^4)$
    \EndIf
    \State Synchronize threads
    \State \textit{// Parallel Spike Compute}
    \For{14 iterations per thread}
        \State $S_{\text{row},j} \leftarrow \round(X_{\text{row},j} / V_{\text{th}}[\text{row}])$
    \EndFor
\EndFor
\State \Return $S, V_{\text{th}}$
\end{algorithmic}
\end{algorithm}

\textbf{CPU-GPU Cooperative Spike Encoding.} Algorithm~\ref{alg:gpu_pulse_encoding} optimizes bitwise spike coding through a hybrid execution model. The GPU first executes Kernel A (Lines 1--2) to compute global extremes via parallel reduction, transferring only two floating-point values to the host to minimize PCIe overhead. The CPU then performs strategy decision (Lines 3--9) in $O(1)$ time to select the optimal transformation mode (e.g., Half-Up), followed by GPU Kernel B (Lines 10--11) which executes in-place coordinate transformation and computes the absolute maximum. After the CPU calculates the bit-width $T$ (Line 12), GPU Kernel C (Lines 13--14) is launched to generate the final bitwise spike sequences. This collaborative design limits host-device communication to 4 floats, bypassing the traditional PCIe bottleneck.

\begin{algorithm}[ht]
\caption{CPU-GPU Cooperative Accelerated Spike Encoding}
\label{alg:gpu_pulse_encoding}
\begin{algorithmic}[1]
\Require Activation $X \in \mathbb{R}^{B \times D}$, flags: $\text{is\_bi}$, $\text{is\_tc}$
\Ensure Bitwise spike sequences $S_{\text{bitwise}} \in \mathbb{R}^{T \times B \times D}$
\State \textbf{GPU (Kernel A):} Compute global $(x_{\min}, x_{\max})$ via parallel reduction
\State Transfer $(x_{\min}, x_{\max})$ from GPU to CPU
\State \textbf{CPU:} Determine encoding $\text{mode}$ based on statistics:
\If {$\text{is\_bi} \land x_{\min} \geq 0$}
    \State $\text{mode} \leftarrow 1$ \Comment{Bidirectional (Half-Up)}
\ElsIf {$\neg\text{is\_bi} \land x_{\min} < 0$}
    \State $\text{mode} \leftarrow 2$ \Comment{Unidirectional (Offset)}
\Else
    \State $\text{mode} \leftarrow 0$ \Comment{No coordinate Transformation}
\EndIf
\State \textbf{GPU (Kernel B):} In-place transform $X$ by $\text{mode}$ and find $\text{abs\_max}$
\State Transfer $\text{abs\_max}$ from GPU to CPU
\State \textbf{CPU:} Calculate bit-width $T \leftarrow \lceil \log_2(\text{abs\_max} + 1) \rceil$
\State \textbf{GPU (Kernel C):} Generate $S_{\text{bitwise}}$ using encoding mode and $T$
\State \Return $S_{\text{bitwise}}, T$
\end{algorithmic}
\end{algorithm}

\textbf{Grouped Symmetric Weight Quantization.} Algorithm~\ref{alg:gpu_weight_quantization} provides a standalone operator for per-channel INT8 quantization using a two-stage parallel pipeline. Stage 1 (Lines 6--9) implements a three-level scale computation: thread-level local max finding, warp-level shuffle reduction, and block-level shared memory aggregation to determine the scaling factor for each group. Stage 2 (Lines 10--18) performs vectorized quantization, where scales are prefetched into shared memory to support irregular indexing (Line 13). By utilizing \texttt{float4} vectorized access (Lines 14 and 17) and handling boundaries with scalar loops, the algorithm achieves near-peak global memory bandwidth while maintaining high numerical precision.

\begin{algorithm}[ht]
\caption{GPU Per-Channel Grouped Symmetric Quantization}
\label{alg:gpu_weight_quantization}
\begin{algorithmic}[1]
\Require Weight matrix $W \in \mathbb{R}^{\text{out\_f} \times \text{in\_f}}$, group size $g = 128$
\Ensure  Quantized weights $W_q$, scales $S \in \mathbb{R}^{\text{out\_f} \times n_g}$
\State \textit{// Preprocessing}
\State $n_g \leftarrow \lceil \text{in\_f} / g \rceil$
\State Adjust $g_{\text{eff}}$ and $n_{g,\text{eff}}$ for boundary cases
\State \textit{// Stage 1: Three-Level Scale Computation}
\State Grid: $(n_{g,\text{eff}}, \text{out\_f})$, Block: 128 threads
\For{each group in parallel}
    \State Thread-level: each processes 4 elements, stride $128 \times 4$
    \State Warp-level: shuffle reduction, offsets $16 \to 1$
    \State Block-level: Thread 0 $S[\text{row},\text{group}] \leftarrow \text{absmax} / 127.0$
\EndFor
\State \textit{// Stage 2: Five-Stage Vectorized Quantization}
\State Grid: $(\lceil \text{in\_f}/512 \rceil, \text{out\_f})$, Block: 128 threads
\For{each block processing 512 elements}
    \State Prefetch: load scales to shared memory
    \State Vectorized load: read 4 floats via float4
    \State Indexing: compute $g_i \leftarrow (\text{col} + i) / g_{\text{eff}}$
    \State Quantize: $q_i \leftarrow \texttt{round}(w_i / s_i)$
    \State Vectorized store: write via float4, handle boundaries
\EndFor
\State \Return $W_q$, $S$
\end{algorithmic}
\end{algorithm}

\subsection{Implementation Details}

\textbf{Implementation of Adaptive Threshold via Offline Kernel Configuration Tool and Two-Stage Reduction.} To maximize the throughput of the adaptive threshold operator, we implement an offline quantitative scoring tool that identifies the optimal thread-block layout for high-dimensional tensors. Recognizing that embedding dimensions in LLMs (e.g., 3584) remain constant while only token counts vary, we perform this evaluation offline to eliminate runtime decision overhead, allowing the pre-determined optimal configuration to be reused throughout the entire inference process. This tool evaluates configurations based on three core dimensions: load balancing ($Score_L$), resource utilization ($Score_R$), and potential hardware occupancy ($Score_O$). By ensuring the integrality of the workload-to-thread ratio, the tool guarantees that each thread processes an identical number of elements (e.g., 14 elements for 3584-dimensional embeddings), thereby eliminating tail cleanup overhead and warp divergence. The implementation leverages several CUDA techniques to bypass traditional bottlenecks. We employ grid-stride loops combined with \texttt{\#pragma unroll} to facilitate instruction-level parallelism (ILP), allowing the GPU to overlap multiple global memory requests. To ensure memory efficiency, the operator utilizes vectorized 128-bit load instructions to achieve perfectly coalesced memory transactions. During the two-stage reduction, we utilize on-chip shared memory as a high-speed scratchpad, synchronized via \texttt{\_\_syncthreads()} to prevent data hazards. For the final stage of reduction, we implement warp-level shuffle primitives (\texttt{\_\_shfl\_down\_sync}) to aggregate partial sums without the latency of shared memory bank conflicts. Furthermore, the operator adaptively employs 2D thread blocks and dynamic shared memory allocation to process multiple rows in parallel. This approach effectively hides memory access latency by increasing the ratio of active warps per SM, ensuring that threshold updates remain off the critical path even during large-scale inference tasks.

\textbf{Fused Kernels and CPU-GPU Collaborative Strategy for Spike Encoding.} To minimize the latency overhead of bitwise spike generation, we implement a collaborative execution model consisting of three specialized CUDA kernels (A, B, and C) interleaved with lightweight CPU decision logic. This architecture bypasses the traditional PCIe bottleneck by restricting host-device communication to only four floating-point values. Kernel A is designed for high-throughput global statistics computation ($x_{\min}, x_{\max}$). We employ 1D grid-stride loops to ensure the kernel can scale to large activation tensors while maintaining coalesced memory access. To avoid the high latency of frequent global memory synchronization, each thread maintains local extrema in registers during the primary traversal. Only after the local processing is complete does each thread perform a single \texttt{atomicMin} and \texttt{atomicMax} operation to aggregate results. This approach reduces the complexity from $O(N)$ on a serial CPU to $O(N/\text{Threads}) + O(1)$ on the GPU. Kernel B implements a fused operator for in-place coordinate transformation and absolute maximum calculation. Depending on the mode selected by the CPU (e.g., bidirectional half-round-up or unidirectional offset), the kernel transforms the input data and writes the results directly back to global memory (\texttt{d\_x}). Crucially, to maximize effective bandwidth, we perform the transformation and the \textit{absolute max aggregation} in a single pass, avoiding a second memory traversal. This fusion ensures that the GPU remains memory-bound rather than instruction-bound, achieving near-peak throughput. Finally, Kernel C executes the parallel bitwise expansion across a 2D thread grid to generate the spike sequence $S_{\text{bitwise}}$. Each thread handles one element and extracts its bitwise representation from high to low using a bit-mask. The kernel supports multiple encoding modes, including bidirectional two's complement and unidirectional unsigned formats. By utilizing a dynamically calculated time step $T$ (derived from the absolute maximum), Kernel C generates sparse spike sequences with minimal bit-depth, significantly reducing the memory footprint for subsequent neuromorphic inference layers.

\textbf{Adaptive Hierarchical Reduction and Vectorized Pipeline for Weight Quantization.} The weight quantization operator is decoupled into two highly parallel stages: scale computation and fused quantization. To achieve peak memory bandwidth, we implement a five-stage vectorized pipeline that leverages \texttt{float4} instructions to execute 128-bit coalesced memory transactions. For the first stage, we develop a three-level hierarchical reduction architecture (thread, warp, and block levels) to calculate per-group scaling factors. This design utilizes \texttt{\_\_shfl\_down\_sync} warp-shuffle primitives to perform intra-warp data exchange directly through registers, eliminating the synchronization overhead and bank conflicts associated with shared memory. The reduction logic adaptively scales with the group size $g$: for $g=128$, threads complete the local max finding in a single iteration; for $g > 512$, threads execute $\lceil g/512 \rceil$ strided iterations to accumulate local maxima, ensuring balanced workloads across all CUDA cores. Boundary cases are handled by dynamically adjusting the active thread count to prevent resource waste.The second stage performs the element-wise quantization using a 2D grid layout where output channels are fully parallelized along the Y-dimension. To resolve the latency bottlenecks of irregular group indexing, we implement a shared memory caching strategy where scaling factors are prefetched into on-chip memory before the main quantization loop. This reduces scale-access latency by two orders of magnitude compared to global memory transactions. While most weight matrix elements are processed using \texttt{float4} vectorized loads, boundary elements are handled via scalar loops to ensure correctness. The two stages are coordinated using sequential CUDA stream launching to maintain data dependencies without extra synchronization overhead.

\section{Evaluation }
This chapter evaluates the proposed $\text{FlashT2S}$ library. Section 6.1-6.2 describe the experimental setup and methodology. Sections 6.3-6.5 discuss functional correctness, operator-level micro-benchmarks, and real-world scenario evaluations.

\subsection{Experimental Setup}
Experiments are conducted on two platforms (Table~\ref{tab:setup}): Platform 1 (NVIDIA A100-40GB) represents enterprise AI servers; Platform 2 (NVIDIA RTX 4090-24GB) represents high-end consumer systems.




\begin{table}[H] 
\caption{Platform Specifications}
\label{tab:setup}
\footnotesize
\setlength{\tabcolsep}{2pt}
\renewcommand{\arraystretch}{1.08}
\begin{tabular}{@{}>{\raggedright\arraybackslash\bfseries}p{0.08\linewidth}
                    >{\raggedright\arraybackslash}p{0.43\linewidth}
                    >{\raggedright\arraybackslash}p{0.46\linewidth}@{}}
\toprule
& \centering\arraybackslash\textbf{Platform 1} &
  \centering\arraybackslash\textbf{Platform 2} \\
\midrule
CPU &
Xeon Gold 5120 @ 2.20 GHz &
Xeon Silver 4314 @ 2.40 GHz \\
& 28c/56t, HT enabled & 32c/64t, HT enabled \\
& L1: 896K, L2: 28M, L3: 38.5M &
  L1: 1.5M/1M, L2: 40M, L3: 48M \\
& 2 NUMA nodes (14c/node) & 2 NUMA nodes (16c/node) \\
& GCC 9.4.0 (\texttt{-O3}) & GCC 11.4.0 (\texttt{-O3}) \\

\bottomrule
\addlinespace[1pt]
GPU &
\textbf{NVIDIA A100 40GB} &
\textbf{NVIDIA RTX 4090} \\
& Ampere (cc 8.0) & Ada Lovelace (cc 8.9) \\
& 6912 CUDA cores & 16384 $\times$ 2 CUDA cores \\
& 40 GB HBM2 & 24 GB GDDR6X/GPU \\
& CUDA 12.9 & CUDA 12.8 \\

\addlinespace[1pt]
OS &
Ubuntu 20.04.6 LTS &
Ubuntu 22.04 LTS \\
\bottomrule
\end{tabular}
\end{table}

\subsection{Evaluation Methodology}
\textbf{Our framework verifies:} (1) \textbf{Correctness}: Ensuring $\text{FlashT2S}$ operators match native implementations \cite{pan2025spikingbrain} output. (2) \textbf{Efficiency}: Quantifying speedup via micro-benchmarks on A100, including comparisons against \texttt{torch.\allowbreak compile} (GPU) (3) \textbf{Generalization}: Testing end-to-end performance across both platforms in real-world scenarios.

\textbf{Baseline Definition:} We adopt the ``native operator'' baseline from the prior SpikingBrain framework \cite{pan2025spikingbrain}
(Source: \path{SpikingBrain-7B/W8ASpike/quant_linear.py}).
Specifically, we compare against the native C++ implementation from this framework
(which follows the same algorithm as the native Python version and differs only in implementation language),
which better reflects real-world production environments.
Additionally, we compare against the GPU kernel generated by \texttt{torch.\allowbreak compile}
from the native Python implementation.

\textbf{Efficiency and Absolute Performance:} Beyond relative speedup, we evaluate the absolute performance of $\text{FlashT2S}$ to demonstrate its near-optimal execution efficiency. While native operators exhibit significant throughput degradation and latency spikes as the sequence length grows, $\text{FlashT2S}$ maintains stable, microsecond-level execution. Specifically, our operators achieve an absolute latency of \textbf{10--99\,$\mu$s} for large-scale data and \textbf{below 10\,$\mu$s} for small-scale inputs. Considering the intrinsic CUDA kernel launch overhead is approximately \textbf{2\,$\mu$s}, these results indicate that $\text{FlashT2S}$ has minimized computational and memory overhead to a level nearing the hardware's theoretical limits.

\subsection{Correctness and Functionality Verification}
$\text{FlashT2S}$ maintains bit-wise consistency with original implementations. The \textbf{Adaptive Threshold} and \textbf{Weight Quantization} operators achieve zero-error conversion. The correctness of the \textbf{Spike Encoding} operator has been rigorously guaranteed across all nine scenarios (Table~\ref{tab:encoding_strategies}) arising from the combination of encoding modes and coordinate transformations.



\begin{table}[ht]
\centering
\caption{Spike Encoding Strategy}
\label{tab:encoding_strategies}
\small
\renewcommand{\arraystretch}{1.2}
\raggedright
\hyphenation{Bi-di-rec-tio-nal Non-Com-ple-ment Trans-for-ma-tion Co-or-di-nate Uni-di-rec-tio-nal}

\setlength{\tabcolsep}{1.5pt}
\begin{tabular}{@{}>{\raggedright\arraybackslash\bfseries}p{0.15\linewidth}
                    >{\raggedright\arraybackslash}p{0.27\linewidth}
                    >{\raggedright\arraybackslash}p{0.27\linewidth}
                    >{\raggedright\arraybackslash}p{0.25\linewidth}@{}}
\toprule
Sample Type & \textbf{Mode 1:} & \textbf{Mode 2:} & \textbf{Mode 3:} \\
& Bidirectional + Non-Complement & Bidirectional + Two's Complement & Unidirectional \\
\midrule
Positive Sample &
(1) $\checkmark$ + Coordinate Transformation &
(2) $\checkmark$ + Coordinate Transformation &
(3) $\checkmark$ \\
\midrule
Negative Sample &
(4) $\checkmark$ &
(5) $\checkmark$ &
(6) $\checkmark$ + Coordinate Transformation \\
\midrule
Mixed Sample &
(7) $\checkmark$ &
(8) $\checkmark$ &
(9) $\checkmark$ + Coordinate Transformation \\
\bottomrule
\end{tabular}
\end{table}

\begin{figure*}[pos=t]
    \centering
    
    \begin{subfigure}[b]{0.43\textwidth}
        \centering
        \includegraphics[width=\linewidth]{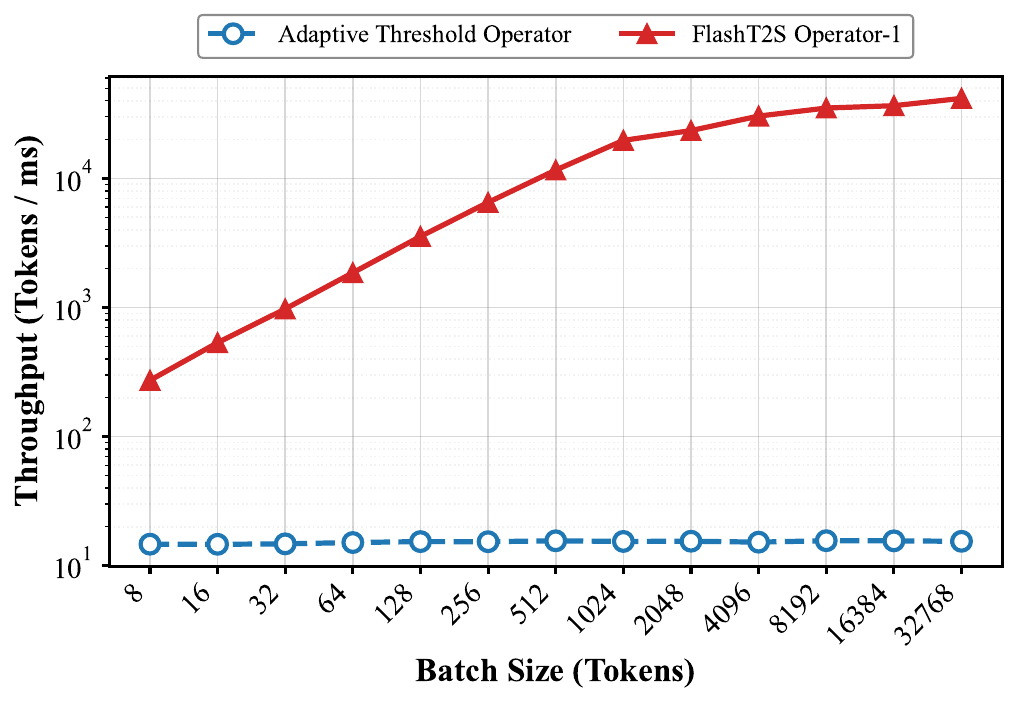}
        \caption{FlashT2S Operator-1 vs Native Adaptive Threshold}
        \label{fig:performance_comparison}
    \end{subfigure}
    \hspace{0.058\textwidth}%
    \begin{subfigure}[b]{0.43\textwidth}
        \centering
        \includegraphics[width=\linewidth]{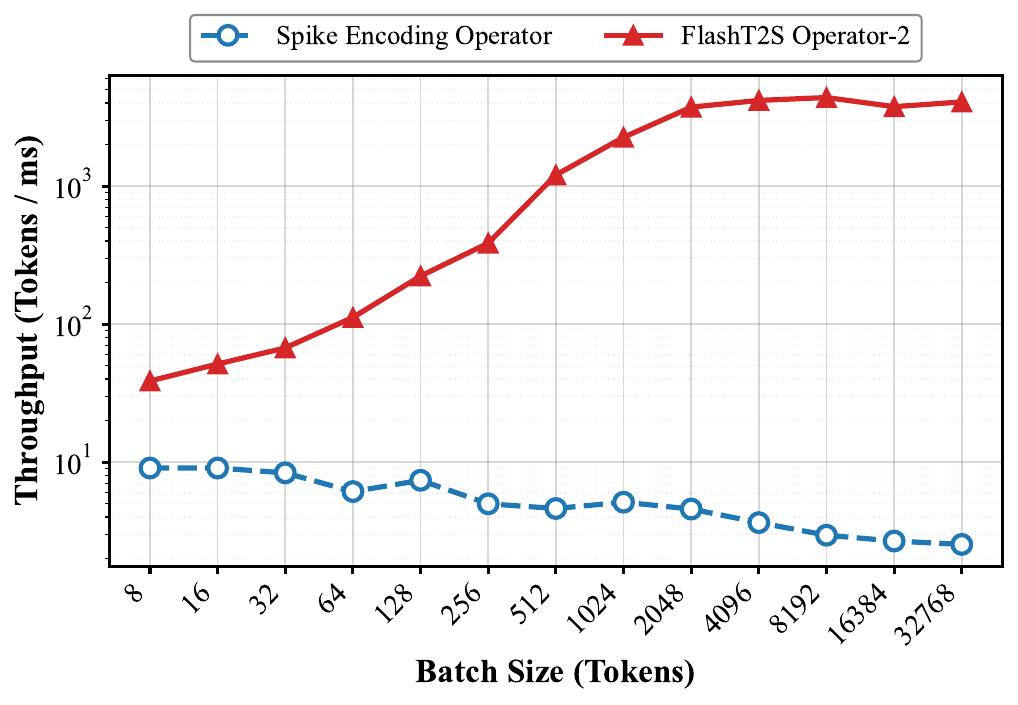}
        \caption{FlashT2S Operator-2 vs Native Spike Encoding}
        \label{fig:spike_encoding_throughput}
    \end{subfigure}

    \caption{Performance comparison across different data scales (Token Hidden Dimension = 3584).}
    \label{fig:combined_performance}
\end{figure*}

\begin{figure*}[pos=t]
    \centering
    \hspace*{23.5pt}%
    \begin{subfigure}[b]{0.42\textwidth}
        \centering
        \includegraphics[width=\linewidth]{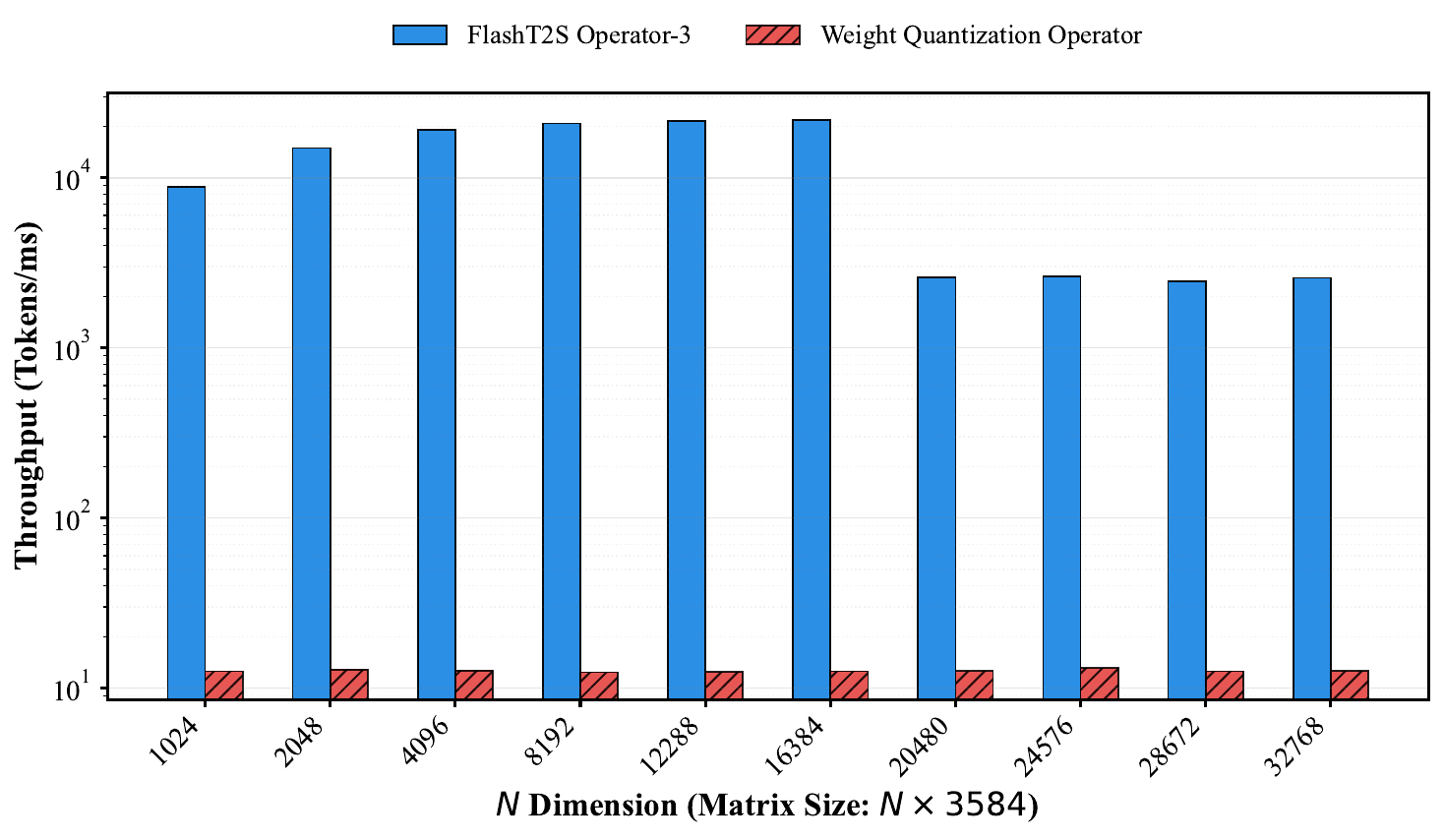}
        \caption{FlashT2S Operator-3 vs Native Weight Quantization (Different Scales)}
        \label{fig:weight_quantization_throughput}
    \end{subfigure}
    \hfill
     \hspace*{18pt}%
    \begin{subfigure}[b]{0.48\textwidth}
        \centering
        \includegraphics[width=\linewidth]{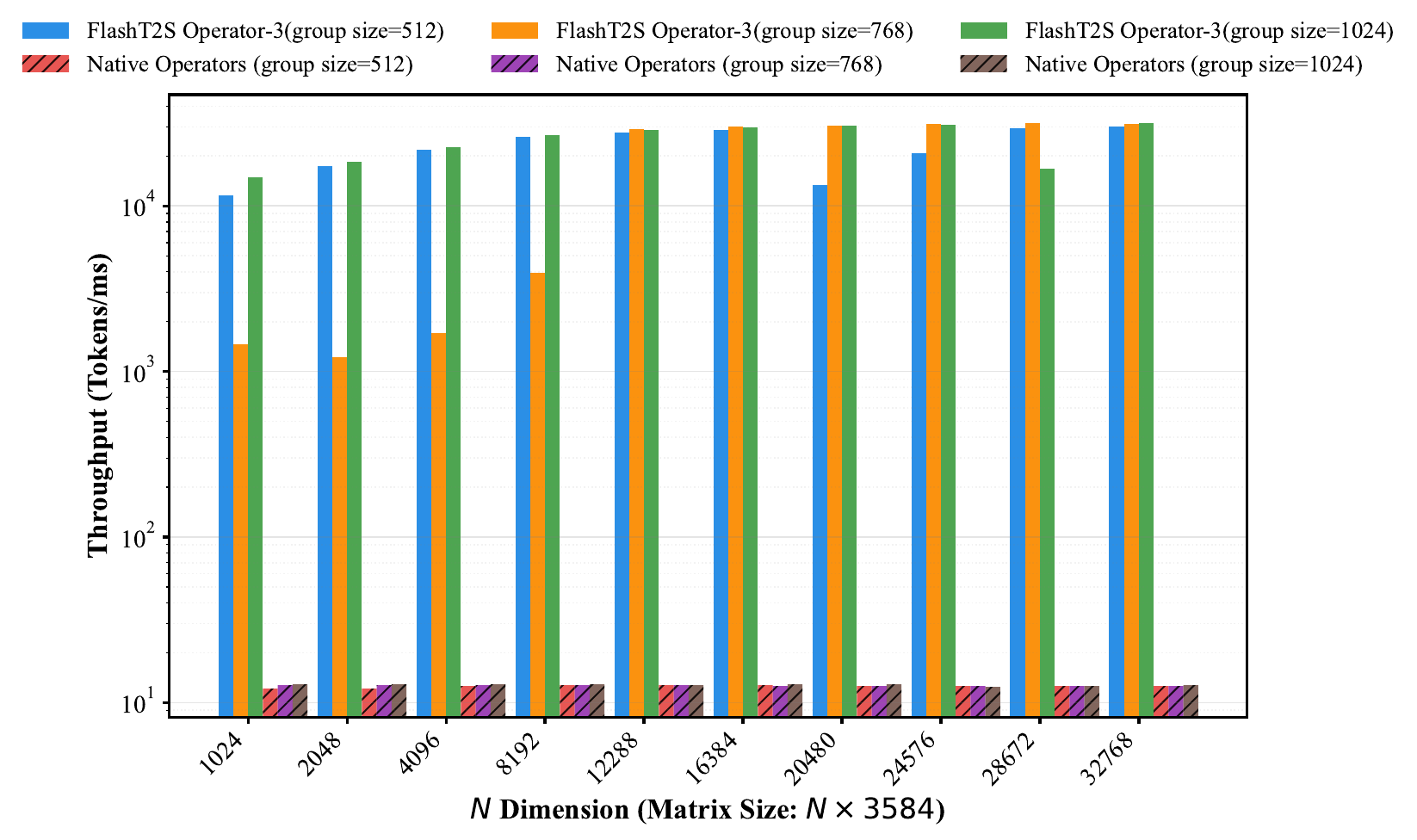}
        \caption{FlashT2S Operator-3 vs Native Weight Quantization (Different Group Sizes)}
        \label{fig:weight_quantization_throughput_grouped}
    \end{subfigure}

    \caption{Performance comparison of parallel-accelerated weight quantization across different scales and group sizes.}
    \label{fig:weight_quantization_combined}
\end{figure*}

\begin{figure*}[pos=t]
    \centering

    \begin{subfigure}[b]{0.44\textwidth}
        \centering
        \includegraphics[width=\linewidth]{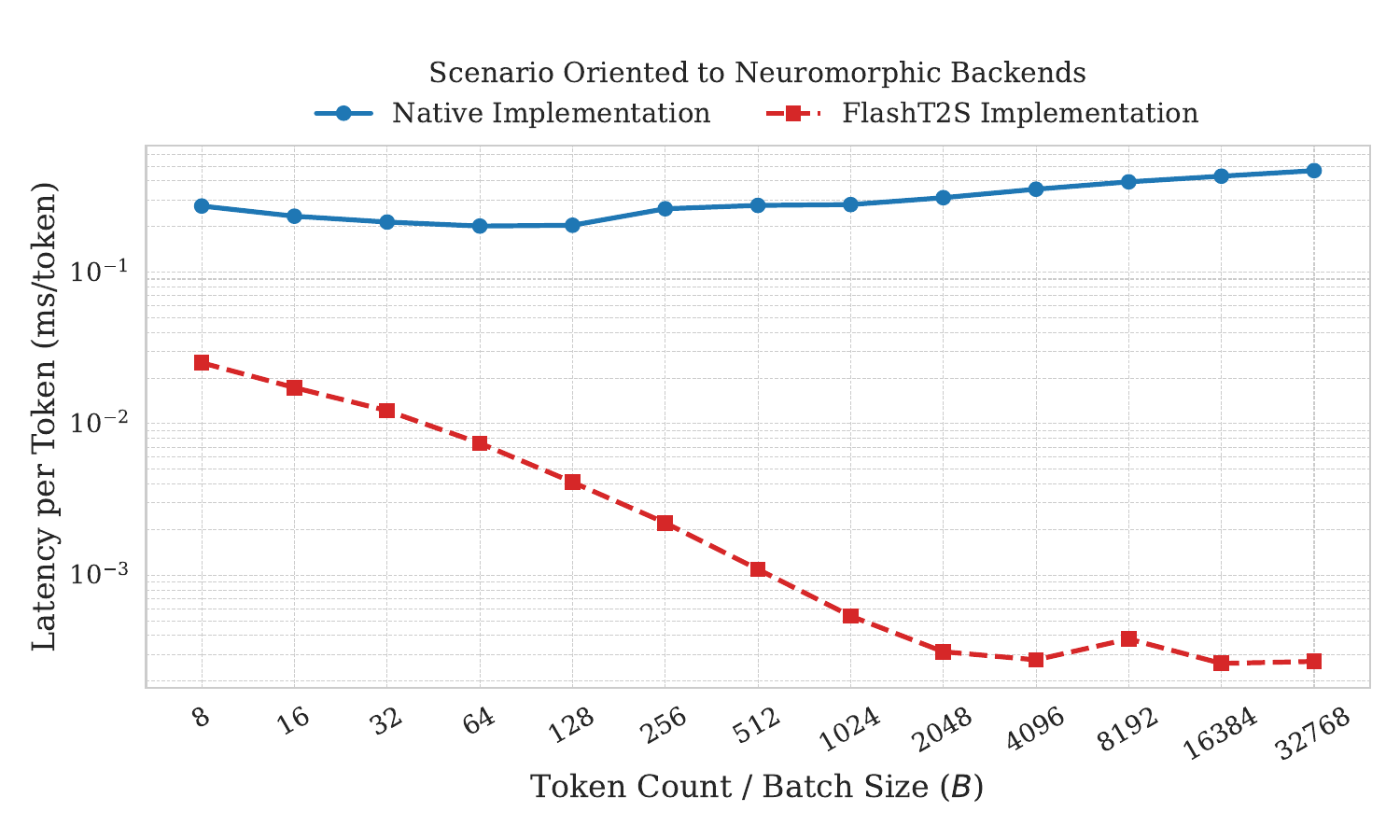}
        \caption{Platform 1 (A100): Neuromorphic Backend-Oriented}
        \label{fig:a100_neuromorphic}
    \end{subfigure}
    \hfill
    \begin{subfigure}[b]{0.44\textwidth}
        \centering
        \includegraphics[width=\linewidth]{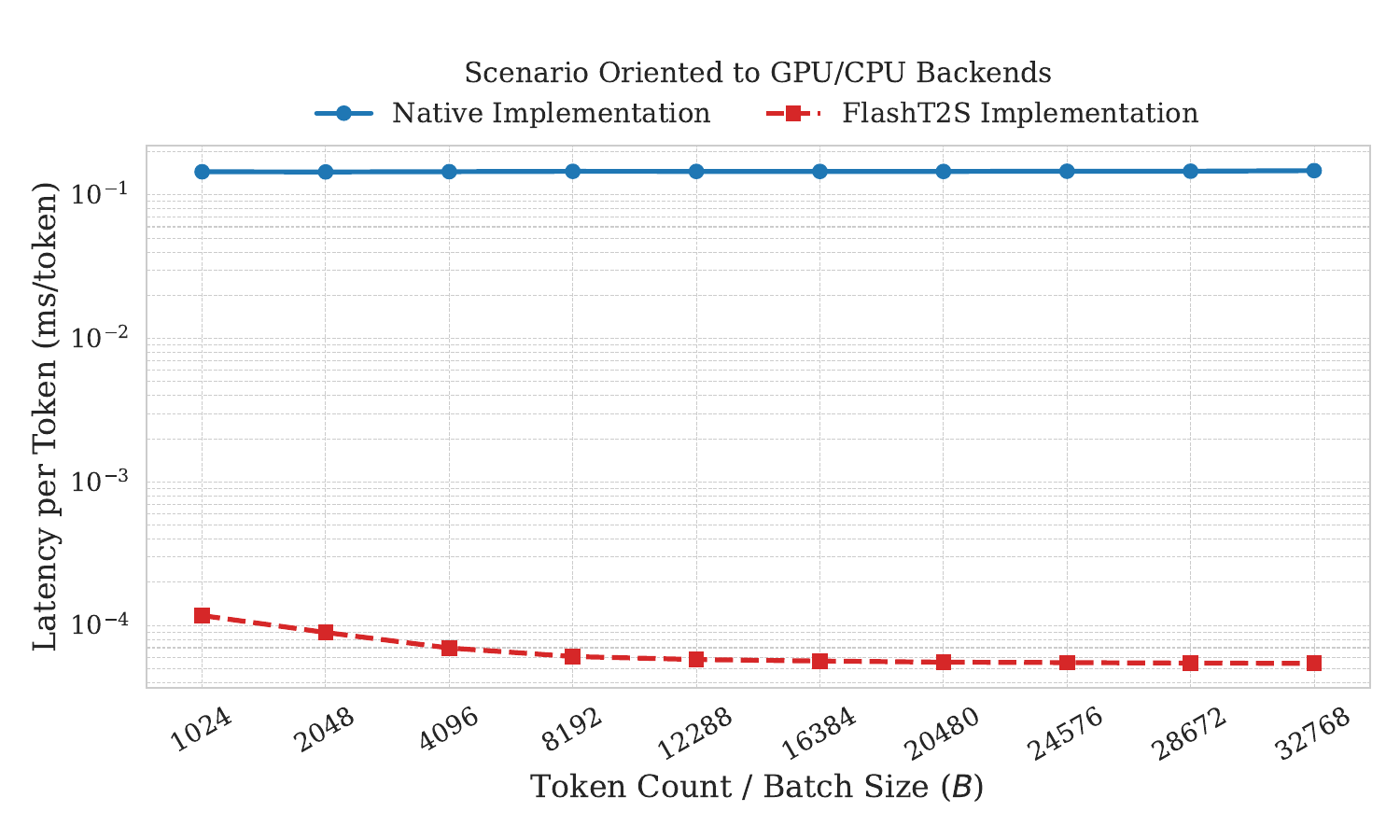}
        \caption{Platform 1 (A100): GPU / CPU Backend-Oriented}
        \label{fig:a100_gpu}
    \end{subfigure}

    \vspace{0.3cm}

    \begin{subfigure}[b]{0.44\textwidth}
        \centering
        \includegraphics[width=\linewidth]{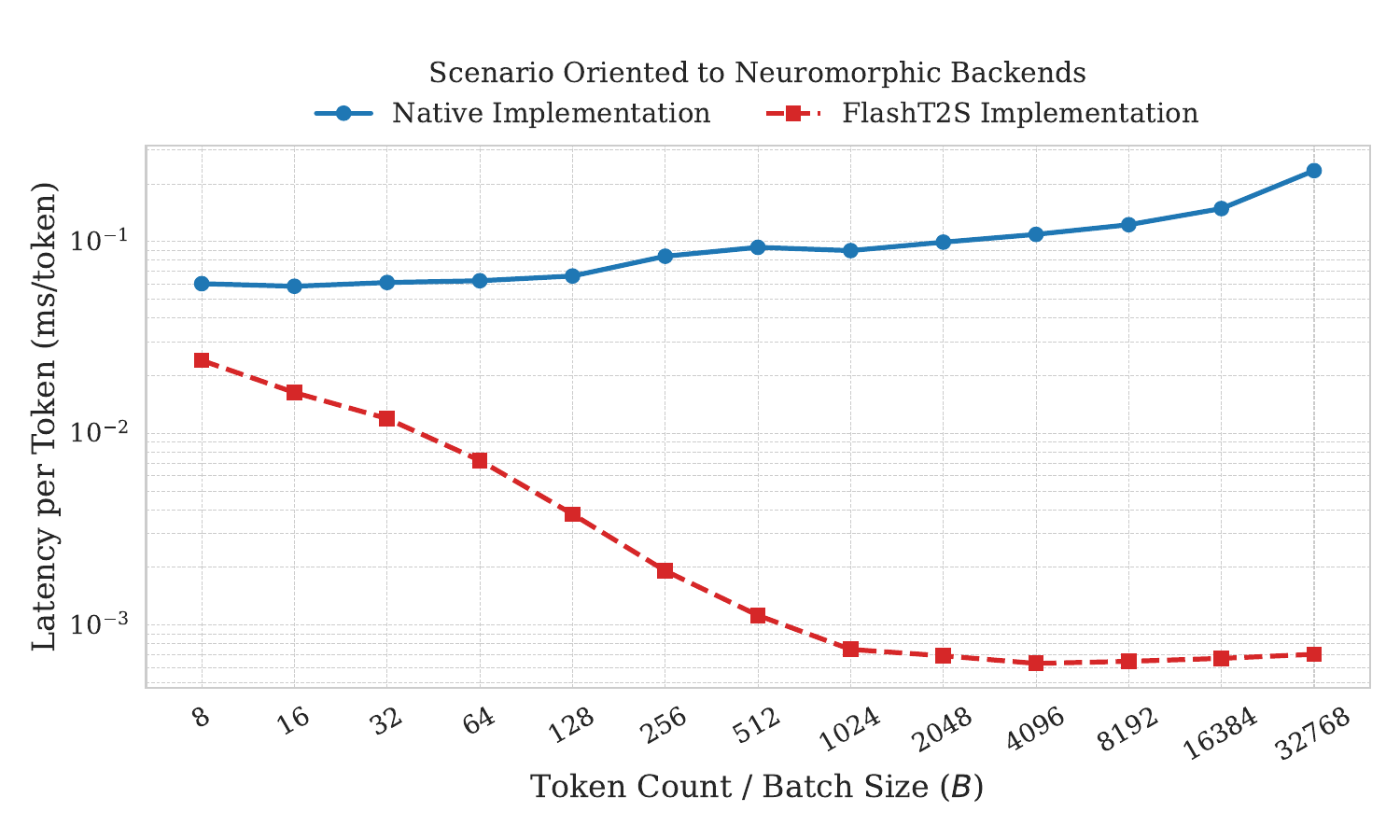}
        \caption{Platform 2 (RTX 4090): Neuromorphic Backend-Oriented}
        \label{fig:4090_neuromorphic}
    \end{subfigure}
    \hfill
    \begin{subfigure}[b]{0.44\textwidth}
        \centering
        \includegraphics[width=\linewidth]{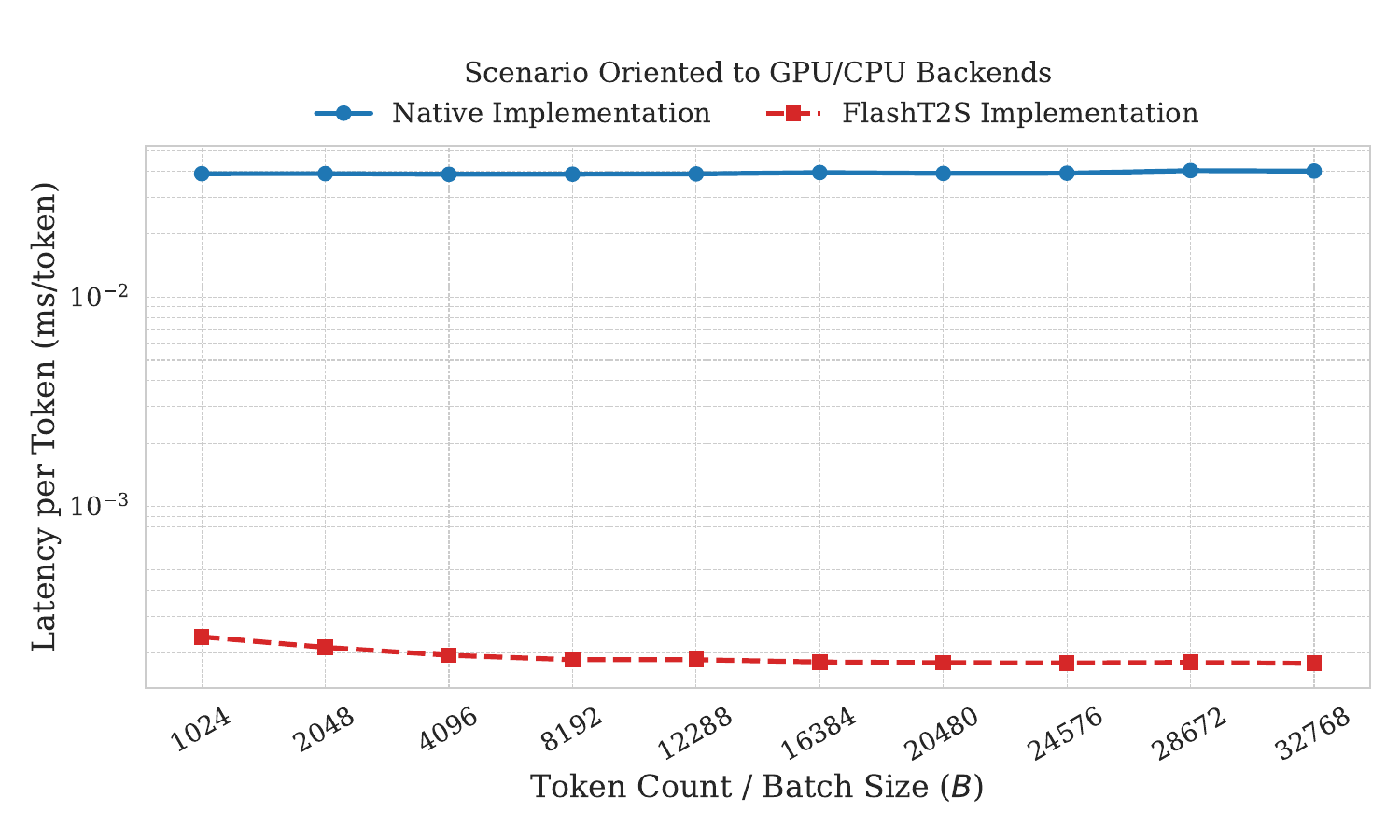}
        \caption{Platform 2 (RTX 4090): GPU / CPU Backend-Oriented}
        \label{fig:4090_gpu}
    \end{subfigure}

    \caption{Latency per token comparison: Native vs. FlashT2S implementations across Platform 1 (NVIDIA A100) and Platform 2 (NVIDIA RTX 4090) for both Neuromorphic and GPU/CPU backend scenarios.}
    \label{fig:combined_latency_comparison}
\end{figure*}
\subsection{Module-Level Performance Evaluation}

\subsubsection{CUDA Core-based Adaptive Threshold Aware Algorithm (Adaptive Threshold Operator)}\mbox{}\

We compare the performance of FlashT2S Operator-1 against the native C++ baseline across sequence lengths from 8 to 32,768 tokens. As shown in Fig. 6(a), FlashT2S achieves significant speedups at all scales: starting from 18.58$\times$ at 8 tokens, rising to 1278.29$\times$ for medium batches (1024 tokens), and reaching a peak of 2698.12$\times$ at 32,768 tokens. 

The core bottleneck of this operator is the two-pass global memory dependency: computing $V_{\mathrm{th}}$ requires a full $\mathcal{O}(D)$ sequential reduction over each row before any spike can be written, forcing the baseline to traverse global memory twice per row. Targeting this bottleneck, Algorithm~\ref{alg:gpu_adaptive_threshold} employs three key optimizations: (1) two-stage parallel reduction to minimize global memory access; (2) shared memory tree reduction for high on-chip throughput; and (3) adaptive grid configuration for optimal thread allocation. These optimizations prove highly effective, reducing latency from milliseconds to the microsecond level and yielding an average speedup of \textbf{1056.12$\times$}.

\subsubsection{GPU-Accelerated Spike Encoding (Spike Bitwise Encoding Operator)}\mbox{}\

FlashT2S Operator-2 is evaluated against the native spike encoding operator (Fig. 6(b)). The core bottleneck of this operator lies in its sequential CPU-side dependency chain: the baseline must complete a full $\mathcal{O}(N)$ extrema search, transfer results to the CPU for mode selection and bit-width calculation, and then re-read the entire activation tensor for encoding---resulting in multiple global memory passes and significant PCIe transfer overhead. To address this, Algorithm~\ref{alg:gpu_pulse_encoding} adopts a CPU-GPU collaborative strategy with three fused CUDA kernels. Kernel A computes global $(x_{\min}, x_{\max})$ via grid-stride loops and register-local extrema aggregated by a single \texttt{atomicMin}/\texttt{atomicMax}, reducing complexity from $\mathcal{O}(N)$ to $\mathcal{O}(N/\text{Threads})$. The CPU then determines the encoding mode and bit-width $T$ from only four transferred floating-point values, keeping PCIe traffic minimal. Kernel B fuses coordinate transformation and absolute maximum aggregation into a single memory pass, and Kernel C performs parallel bitwise expansion over a 2D thread grid to generate the final sparse spike sequence $S_{\text{bitwise}}$. As shown in Fig.~6(b), these optimizations prove highly effective: speedup scales from $4.28\times$ at 8 tokens to $1611.90\times$ at 32,768 tokens, achieving an average speedup of $\textbf{563.07}\times$.

\subsubsection{Parallel-Accelerated Weight Quantization (Weight Quantization Operator)}\mbox{}\

We evaluate FlashT2S Operator-3 across various matrix dimensions and group sizes (Fig.~\ref{fig:weight_quantization_combined}). The core bottleneck of grouped symmetric quantization is its two-pass dependency: scale computation requires a full per-group $\mathcal{O}(g)$ reduction before any element can be quantized, and irregular group indexing during the second pass introduces non-coalesced memory access that further degrades throughput. To address this, Algorithm~\ref{alg:gpu_weight_quantization} decouples the operator into two stages. Stage 1 resolves per-group scales via a three-level hierarchical reduction (thread $\to$ warp $\to$ block), where warp-shuffle primitives (\texttt{\_\_shfl\_down\_sync}) eliminate shared memory overhead and the design adapts to arbitrary group sizes through strided iterations. Stage 2 prefetches scales into shared memory and applies a five-stage vectorized pipeline with \texttt{float4} 128-bit loads, converting irregular group indexing from a global memory bottleneck into an on-chip lookup.

As shown in Fig.~\ref{fig:weight_quantization_combined}(a), these optimizations yield speedups rising from $705.82\times$ at dimension 1,024 to a peak of $1735.26\times$ at 16,384, with an average of $\textbf{933.16}\times$. The monotonic growth with matrix dimension confirms that the vectorized pipeline increasingly amortizes launch overhead as problem size grows. Fig.~\ref{fig:weight_quantization_combined}(b) further shows that speedup remains stable across group sizes 512, 768, and 1024, validating that the adaptive strided reduction and shared memory broadcast together absorb the irregular indexing cost regardless of group boundary alignment.

\subsubsection{Comparison of Three Operators with \texttt{torch.\allowbreak compile}-Generated GPU Kernels}\mbox{}\

We further compile the native operators with \texttt{torch.\allowbreak compile} to generate GPU kernels and compare them with our manually designed parallel algorithms. As shown in Tables~\ref{tab:triton}--\ref{tab:quant_torch_compile}, FlashT2S achieve average speedups of \textbf{7.21$\times$}, \textbf{11.47$\times$}, and \textbf{5.01$\times$} for adaptive thresholding, spike encoding, and weight quantization, respectively.

For Operator-1 (Table~\ref{tab:triton}), \texttt{torch.\allowbreak compile} performs reasonably well at large sequence lengths, narrowing the gap to $2.20\times$--$2.55\times$ beyond 8,192 tokens, as the row-wise reduction structure is relatively amenable to auto-generated tiling. However, our hand-tuned two-stage kernel retains a substantial advantage at small batches ($12.84\times$ at 8 tokens), where adaptive grid switching and shared-memory caching provide occupancy benefits that \url{torch.compile} cannot recover through generic fusion.

\begin{table}[ht]
  \caption{Speedup of FlashT2S Operator-1 over \texttt{torch.\allowbreak compile} GPU kernel.}
  \label{tab:triton}
  \centering
  \small
  \setlength{\tabcolsep}{3.6pt}
  \renewcommand{\arraystretch}{1.15}

  \resizebox{0.95\linewidth}{!}{%
  \begin{tabular}{@{}lccccccc@{}}
    \toprule
    \textbf{Seq. Len.} & 8 & 16 & 32 & 64 & 128 & 256 & 512 \\
    \midrule
    \textbf{Speedup} &
    12.84$\times$ & 12.95$\times$ & 10.31$\times$ & 9.80$\times$ &
    10.06$\times$ & 9.07$\times$ & 8.06$\times$ \\
    \midrule
    \textbf{Seq. Len.} & 1024 & 2048 & 4096 & 8192 & 16384 & 32768 & \textbf{Avg.} \\
    \midrule
    \textbf{Speedup} &
    6.85$\times$ & 3.71$\times$ & 3.04$\times$ & 2.55$\times$ &
    2.20$\times$ & 2.36$\times$ & \textbf{7.21$\times$} \\
    \bottomrule
  \end{tabular}}
\end{table}

For Operator-2 (Table~\ref{tab:encode_torch_compile}), the gap widens at larger sequence lengths rather than narrowing, reaching $16.87\times$--$17.32\times$ beyond 2,048 tokens. This reflects the fundamental limitation of \url{torch.compile} on multi-stage pipelines with CPU-GPU control flow: the auto-compiler cannot fuse across the CPU decision steps (mode selection, bit-width calculation) that sit between our CUDA kernels, whereas our collaborative architecture explicitly minimizes PCIe transfers to four floating-point values and keeps all heavy computation on-device.

\begin{table}[ht]
  \caption{Speedup of FlashT2S Operator-2 over \texttt{torch.\allowbreak compile} GPU kernel.}
  \label{tab:encode_torch_compile}
  \centering
  \small
  \setlength{\tabcolsep}{3.6pt}
  \renewcommand{\arraystretch}{1.15}

  \resizebox{0.95\linewidth}{!}{%
  \begin{tabular}{@{}lccccccc@{}}
    \toprule
    \textbf{Seq. Len.} & 8 & 16 & 32 & 64 & 128 & 256 & 512 \\
    \midrule
    \textbf{Speedup} &
    12.47$\times$ & 8.37$\times$ & 6.14$\times$ & 5.53$\times$ &
    6.10$\times$ & 6.00$\times$ & 11.35$\times$ \\
    \midrule
    \textbf{Seq. Len.} & 1024 & 2048 & 4096 & 8192 & 16384 & 32768 & \textbf{Avg.} \\
    \midrule
    \textbf{Speedup} &
    13.52$\times$ & 16.87$\times$ & 17.32$\times$ & 14.41$\times$ &
    14.69$\times$ & 16.40$\times$ & \textbf{11.47$\times$} \\
    \bottomrule
  \end{tabular}}
\end{table}

For Operator-3 (Table~\ref{tab:quant_torch_compile}), our kernel leads consistently up to 16,384 dimensions but falls below $1\times$ beyond 20,480, where \url{torch.compile} benefits from its generic tiling strategy on very large regular matrices. This indicates that our vectorized pipeline and hierarchical reduction are most advantageous when group boundary irregularities dominate; at extreme dimensions the workload becomes sufficiently regular that auto-generated kernels can close the gap. Overall, these results confirm that operator-specific designs are most impactful precisely where \texttt{torch.\allowbreak compile} struggles: irregular data dependencies, multi-stage control flow, and small-to-medium batch regimes common in neuromorphic inference.

\begin{table}[ht]
  \caption{Speedup of FlashT2S Operator-3 over \texttt{torch.\allowbreak compile} GPU kernel.}
  \label{tab:quant_torch_compile}
  \centering
  \small
  \setlength{\tabcolsep}{3.6pt}
  \renewcommand{\arraystretch}{1.15}

  \resizebox{0.95\linewidth}{!}{%
  \begin{tabular}{@{}lcccccc@{}}
    \toprule
    \textbf{Seq.} & 1024 & 2048 & 4096 & 8192 & 12288 & \textbf{Avg.} \\
    \midrule
    \textbf{Speedup} &
    3.81$\times$ & 14.34$\times$ & 6.84$\times$ & 7.03$\times$ &
    7.48$\times$ & \textbf{5.02$\times$} \\
    \midrule
    \textbf{Seq.} & 16384 & 20480 & 24576 & 28672 & 32768 &  \\
    \midrule
    \textbf{Speedup} &
    7.23$\times$ & 0.86$\times$ & 0.87$\times$ & 0.81$\times$ &
    0.87$\times$ &  \\
    \bottomrule
  \end{tabular}}
\end{table}

\subsection{Real-World Scenario Evaluation}

Real-world scenario performance is tested on Platform 1 (A100) and Platform 2 (RTX 4090) for two critical inference paths: neuromorphic backends (Op-1 $+$ Op-2) and GPU/CPU backends (Op-1 $+$ Op-3).

As shown in Fig.~\ref{fig:combined_latency_comparison}, the latency-per-token curves reveal a consistent pattern across both platforms: the native implementation grows with batch size due to its sequential computation, while FlashT2S maintains a steeply declining per-token latency, reflecting the amortization of fixed GPU launch overhead over larger batches. On Platform 1 (A100), FlashT2S achieves an average speedup of $591.10\times$ for neuromorphic scenarios (Fig.~\ref{fig:combined_latency_comparison}(a)) and $2311.42\times$ for GPU/CPU hybrid scenarios (Fig.~\ref{fig:combined_latency_comparison}(b)). The substantially larger gain in the GPU/CPU path is expected: Op-3's vectorized pipeline and hierarchical reduction exploit the A100's high-bandwidth memory more aggressively than the spike encoding path, whose CPU-GPU control flow introduces a small but non-negligible synchronization cost. On Platform 2 (RTX 4090), FlashT2S maintains a significant advantage with average speedups of $103.51\times$ (Fig.~\ref{fig:combined_latency_comparison}(c)) and $205.29\times$ (Fig.~\ref{fig:combined_latency_comparison}(d)), respectively. The lower absolute speedups relative to the A100 are consistent with the RTX 4090's narrower memory bandwidth and smaller shared memory capacity, which moderately limit the benefit of our on-chip reduction strategies---yet the gap remains over two orders of magnitude, confirming that our optimizations are not hardware-specific.

In summary, across both enterprise-grade (A100) and consumer-grade (RTX 4090) hardware, FlashT2S delivers an overall average speedup of $\textbf{802.83}\times$, reducing data conversion latency from hundreds of milliseconds to the microsecond level. This makes real-time inference and high-throughput deployment of neuromorphic LLMs practical across a broad range of hardware configurations.

\section{Related Work }
\textbf{Efficient LLMs and Quantization.}
To mitigate the ``memory wall'' in Large Language Models (LLMs) \cite{gholami2024ai,brown2020language}, research has focused on sparsity \cite{frantar2023sparsegpt}, LoRA \cite{hu2022lora}, and quantization. Mainstream methods like SmoothQuant \cite{xiao2023smoothquant}, AWQ \cite{lin2024awq}, and GPTQ \cite{frantar2022gptq} optimize dense GEMM kernels for lower bit-widths. On the system level, FlashAttention \cite{dao2022flashattention,dao2023flashattention} reduces memory overhead through tiling. However, these frameworks, including QLoRA \cite{dettmers2023qlora}, are designed for continuous tensors and lack native support for discrete, asynchronous operators like spike encoding. In contrast, FlashT2S is designed to bridge the gap between dense tensors and neuromorphic-style computation.

\textbf{Neuromorphic Computing and Spiking LLMs.} 
Spiking Neural Networks (SNNs) provide an energy-efficient alternative via event-driven computation \cite{roy2019towards}. While hardware like Loihi \cite{davies2018loihi} and algorithms like surrogate gradients \cite{neftci2019surrogate} enabled deep SNNs, recent models such as SpikeGPT \cite{zhu2023spikegpt} and Spikeformer \cite{li2024spikeformer,yao2023spike,peng2023rwkv} demonstrate their potential for LLMs. Despite this, a significant gap exists between GPU-based training and neuromorphic inference \cite{schuman2022opportunities}. Current hybrid systems \cite{pan2025spikingbrain,khadse2025neuromorphic} suffer from inefficient dataflow conversion between tensors and spikes. Our work introduces efficient conversion operators to eliminate these system-level bottlenecks.

\textbf{ANN-to-SNN Conversion.} 
ANN-to-SNN conversion maps pre-trained ANN weights to spiking neurons. Early rate-coding methods \cite{rueckauer2017conversion} often faced high latency \cite{bu2023optimal}. Recent advances have optimized this via layer-wise distillation \cite{hong2025lasnn}, negative spike compensation \cite{wang2025fast}, and precise neuron fitting \cite{wang2024universal,wang2025precise}. Frameworks like NeuBridge \cite{yang2025neubridge} and FAS \cite{chen2025fas} focus on bridging quantized activations with spiking neurons. Unlike prior work focused on algorithmic mapping, FlashT2S leverages GPU parallelism to enable high-performance spike conversion for diverse neuromorphic backends.

\section{Conclusion }
This paper introduces \textbf{FlashT2S}, a library of high-performance GPU-accelerated operators designed for heterogeneous neuromorphic LLM backends, effectively addressing the critical data conversion bottleneck. Our empirical analysis of hybrid heterogeneous systems reveals a critical \textbf{mismatch}: while downstream computation is inherently fast, the prerequisite tensor-to-spike conversion often becomes the primary bottleneck. 

To bridge this gap, we proposed architecture-aware and data-centric parallel algorithms that exploit fine-grained GPU parallelism for multi-step complex operators, including adaptive thresholding, spike bitwise encoding, and parallel weight quantization. Experimental results demonstrate that FlashT2S achieves a significant leap in performance, with independent operators reaching an average speedup of \textbf{850.78$\times$} and realistic workloads showing an average speedup of \textbf{802.83$\times$}. Moreover, FlashT2S achieves \textbf{5--11$\times$} speedups over GPU kernels automatically generated by \texttt{torch.\allowbreak compile}.

Notably, FlashT2S optimizes conversion efficiency to near-optimal levels: large-scale tensor-to-spike conversions are reduced to tens of microseconds \textbf{(10--99 $\mu$s)}, while small-scale conversions reach single-digit microsecond levels \textbf{($<$10 $\mu$s)}. By transforming conversion latency from a millisecond-scale hurdle into a microsecond-scale negligible cost, FlashT2S effectively eliminates the communication overhead between von Neumann and neuromorphic architectures. This work not only provides an immediate performance boost for hybrid LLMs but also establishes a scalable framework for efficient data synchronization in the evolving neuromorphic computing ecosystem.


\section*{Declaration of competing interest}
The authors declare that they have no known competing financial interests or personal relationships that could have appeared to influence the work reported in this paper.

\section*{Data availability}
The source code is publicly available in the Gitee repository at \url{https://gitee.com/GuoEulerLab/SpikeTransOps} for use by potential users.

\bibliographystyle{cas-model2-names}
\bibliography{references}

\end{document}